\documentclass[]{aastex701}
\usepackage{amsmath}
\usepackage{threeparttable}

\begin{document}

\title{Gemini and Apache Point Multi-band Optical Imaging Characterization of Fragmenting Long-period Comet C/2025 K1 (ATLAS)}

\author[orcid=0000-0002-7053-5495]{Carl Ingebretsen}
\affiliation{William H. Miller III Department of Physics and Astronomy, Johns Hopkins University, 3400 N. Charles Street, Baltimore, MD 21218, USA}
\email{cingebr1@jh.edu}

\author[orcid=0000-0002-4950-6323]{Bryce T. Bolin}
\affiliation{Eureka Scientific, Oakland, CA 94602, USA}
\email{bolin.astro@gmail.com} 

\author[orcid=0000-0001-7891-8143]{Meredith A. MacGregor}
\affiliation{William H. Miller III Department of Physics and Astronomy, Johns Hopkins University, 3400 N. Charles Street, Baltimore, MD 21218, USA}
\email{mmacgregor@jhu.edu}

\author[orcid=0000-0002-9548-1526]{Carey~M.~Lisse}
\affiliation{Johns Hopkins University Applied Physics Laboratory, Planetary Exploration Group, Space Department, 11100 Johns Hopkins Road, Laurel, MD 20723, USA}
\email{carey.lisse@jhuapl.edu}

\author[orcid=0000-0003-4778-6170]{Matthew Belyakov}
\affiliation{Division of Geological and Planetary Sciences, California Institute of Technology, Pasadena, CA 91125, USA}
\email{mattbel@caltech.edu}

\author[orcid=0009-0009-9105-7865]{Gracyn Jewett}
\affiliation{ Department of Physics and Astronomy, University of Oklahoma, Norman, OK 73019, USA}
\email{gjewett@ou.edu}

\author[orcid=0000-0001-6098-2235]{Mukremin Kilic}
\affiliation{ Department of Physics and Astronomy, University of Oklahoma, Norman, OK 73019, USA}
\email{kilic@ou.edu}

\author[orcid=0000-0001-7895-8209]{Marco Micheli}
\affiliation{European Space Agency, Near Earth Object Coordination Centre, Frascati, Italy}
\email{marco.micheli@esa.int}

\author[orcid=0000-0003-0774-884X]{Davide Farnocchia}
\affiliation{Jet Propulsion Laboratory, California Institute of Technology, Pasadena, CA 91109, USA}
\email{davide.farnocchia@jpl.nasa.gov}

\author[orcid=0000-0002-1428-7036]{Brian Lemaux}
\affiliation{International Gemini Observatory/NSF NOIRLab, Hilo, HI 96720, USA}
\email{brian.lemaux@noirlab.edu}

\author[orcid=0000-0002-2536-1633]{Hyewon Suh}
\affiliation{International Gemini Observatory/NSF NOIRLab, Hilo, HI 96720, USA}
\email{hyewon.suh@noirlab.edu}

\begin{abstract}

We present optical imaging of the fragmented long-period comet C/2025 K1 (ATLAS) obtained with Gemini North/GMOS and the Apache Point Observatory 3.5-m (ARC 3.5-m) telescope/ARCTIC on 2025 December 6, 8, and 24. Gemini r-band imaging on December 6 revealed four fragments, designated A, D, E, and C. The December 8 ARC 3.5-m u, g, r, i, and z imaging resolved C and a blended A+D fragment, while Gemini g, r, and i imaging on December 24 resolved the A, D, and C fragments. The E fragment was not recovered. Approximate changes in the projected A-D and A-C separations correspond to projected plane-of-sky separation rates of approximately 1 ${\rm m\ s^{-1}}$ and 3 ${\rm m \ s^{-1}}$. The measured broadband coma colors span g-r$\simeq$0.35 and r-i$\simeq$0.2. No significant color change is detected for C between December 8 and 24. Approximately matched-aperture r-band measurements indicate fading of C by December 24. Together, these observations provide a temporally resolved view of the survival, dispersal, and photometric evolution of C/2025 K1's fragment system during the weeks following its breakup in early 2025 November.
\end{abstract}

\section{Introduction}
\label{sec:intro}

Dynamically new Oort Cloud comet C/2025 K1 (ATLAS) was first discovered in May 2025 by the Asteroid Terrestrial-impact Last Alert System \citep[ATLAS,][]{Tonry2018,Denneau2025K1MPEC}. Orbital calculations indicate that it originated from the Oort Cloud, with an eccentricity of $\sim$ 1.001, and reached a perihelion point only 0.33~au from the Sun on October 8, 2025.  Following this close approach, the comet experienced a major activity increase between November 2-4, 2025, along with changes in coma morphology \citep{Bodewits2026K1} before fragmenting into two components \citep{serra-ricart:2025,Kostov2025K1Fragment}.  Deeper follow-up observations with HST showed that there were in fact five fragments \citep{Bodewits2026K1, noonan:2025}.  As it moved farther from the Sun, the comet continued to fragment with secondary fragmentation of the C fragment reported on January 10, 2026 in JWST NIRCam images \citep{bolin:2026}.

Spectroscopic observations in 2025 August before the comet's 2025 October perihelion passage demonstrate that C/2025 K1 is significantly depleted in carbon-chain molecules, including C$_2$ and C$_3$ \citep[][]{manzini:2025}.  Additional spectroscopic measurements were taken on 2025 November 13, shortly after fragmentation, showing the same depletion and significant detections of NH$_2$ \citep[][]{ganesh:2025}. Calculated production rates of NH, C$_2$, C$_3$, and OH are lower than expected for a typical comet at a similar distance ($\sim1.5$~au) \citep{ganesh:2025,Jehin2025K1}. Additional observations support this conclusion, showing very low CN-to-OH and dust-to-gas ratios \citep[][]{schleicher:2025}. Surprisingly, the comet does exhibit strong NH$_2$ emission \citep[][]{ganesh:2025}.  These properties are atypical for solar system comets with only two other comets observed to have such carbon-poor compositions \citep[C/1988 Y1 Yanaka and 96P/Machholz,][]{schleicher:2025}.

In this paper, we present multi-band imaging of Comet C/2025 K1 (ATLAS) using data from the Apache Point Observatory (APO) and Gemini North Observatory. This provides a new data epoch, roughly two months after perihelion and one month after the observed fragmentation. In that time, the object moved from 0.33~au at perihelion to 1.4~au on 2025 December 8 and underwent a major splitting event. We present the observations as obtained in Section~\ref{sec:obs}. Section~\ref{sec:analysis} presents the results of our analysis, including K1's blue g-r color and the evolution of the fragments' dust-scattering coma brightness. We then discuss the implications of our results and conclusions in Section~\ref{sec:discussion} and Section~\ref{sec:conclusions}, respectively.

\section{Observations} 
\label{sec:obs}

We observed fragmenting comet C/2025 K1 with both the Gemini North Observatory 8.1-m telescope and the APO Astrophysical Research Consortium (ARC) 3.5-m telescope.  Additional details on each set of observations are provided below.

\subsection{Gemini North Observatory}
\label{sec:gemini}

\begin{figure}
    \centering
    \includegraphics[width=.55\linewidth]{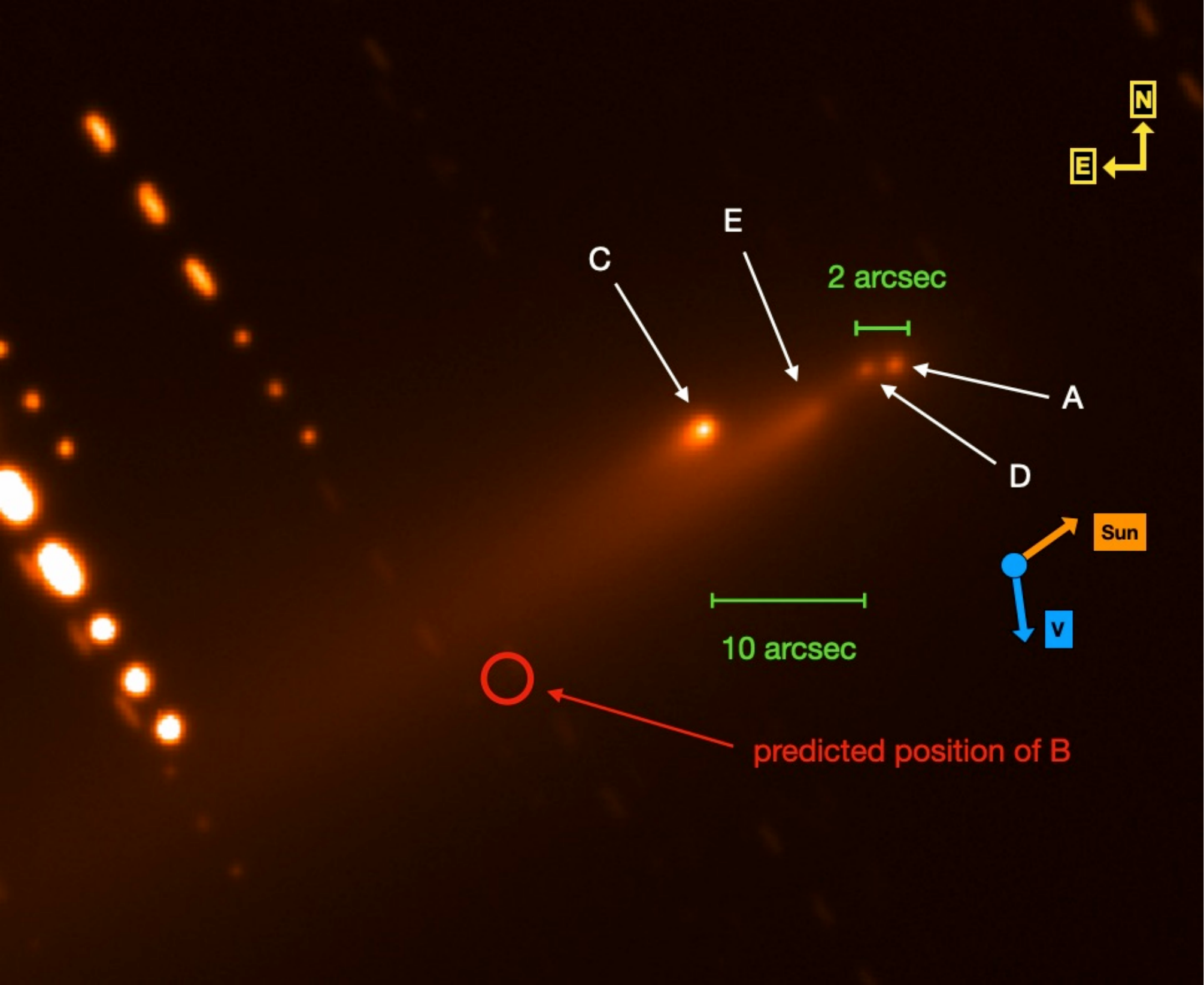}
    \caption{Gemini/GMOS r-band data of C/2025 K1 taken on 2025 December 6. The image consists of six separate images stacked with a combined integration time of 69 s. Individual fragments, cardinal direction, as well as solar direction, heliocentric velocity direction, and image scale are indicated. The locations of the A, D, E, and C fragments are indicated. The predicted position of the B fragment is also indicated.}
    \label{fig:GemDec6}
\end{figure}

We obtained two epochs of imaging of C/2025 K1 with the 8.1-m Gemini North telescope and the Gemini Multi-Object Spectrograph \citep[GMOS,][]{Hook2004}. The images obtained with GMOS used the Hamamatsu in $2\times2$ binning mode, providing a pixel scale of $0\farcs161$, low amplifier gain mode, and slow readout mode. Preliminary accounts of these observations were reported by \citet{bolin_atel1} and \citet{bolin_atel2}. The first epoch was obtained on 2025 December 6 under program GN-2025B-DD-113 (PI: Bolin). The observing sequence extended from 08:47:25 to 08:50:59. The telescope tracked at the non-sidereal rate of the comet of $6\farcs2$~min$^{-1}$, and GMOS acquired 3 exposures of 3~s each and an additional three exposures of 20~s each through the r filter, giving a total integration time of 69~s. The effective wavelength of the r filter was approximately 630~nm. The delivered image quality was approximately $0\farcs7$, and the observations covered an airmass of $\sim$1.44. Table~\ref{tab:obs} summarizes the viewing geometry of the GMOS observations. At the midpoint of the 2025 December 6, sequence, the comet was at heliocentric and geocentric distances of 1.41 and 0.55~au, respectively, and at a phase angle of $32\fdg4$. As seen in Fig.~\ref{fig:GemDec6}, four condensations were visible over a projected extent of approximately 13\arcsec, corresponding to about 5200~km at the comet. Using the nomenclature adopted throughout this paper, these components are labeled A, D, E, and C from west to east. Components A and D were separated by approximately 2\arcsec, $\sim$800~km at the distance of the comet. Fragment E appeared as a more diffuse, dusty condensation between D and C. The fragment trail had a position angle of approximately 110\degr, close to the projected Sun-to-target radius vector angle of $\sim$125\degr. A dust tail extended approximately 90\arcsec, or 36,000~km at the distance of the comet with a position angle of $\sim$120\degr. Component B was not detected at its predicted position; the corresponding detection limit is determined in the Results section below.

\begin{deluxetable*}{llccccccc}
\tablecaption{Summary of Observational Circumstances for C/2025 K1
\label{tab:obs}}
\tablehead{
\colhead{Observation Data} &
\colhead{Facility} &
\colhead{Filter} &
\colhead{$\theta_s^1$} &
\colhead{$\chi_{am}^2$} &
\colhead{$r_H^3$} &
\colhead{$\Delta^4$} &
\colhead{$\alpha^5$} &
\colhead{$\delta_E^6$} \\
\colhead{[UTC]} &
\colhead{} &
\colhead{} &
\colhead{[\arcsec]} &
\colhead{} &
\colhead{[au]} &
\colhead{[au]} &
\colhead{[\degr]} &
\colhead{[\degr]} 
}
\startdata
2025 December 6  & Gemini North & r          & 0.7            & 1.44          & 1.41    & 0.55       & 32.4         & -22.9        \\
2025 December 8  & APO          & u, g, r, i, z & 1.2            & 1.37          & 1.44    & 0.59       & 31.9         & -19.2        \\
2025 December 24 & Gemini North & g, r, i    & 0.6            & 1.12          & 1.72    & 1.06       & 31.5         & -2.8         \\
\enddata
\tablecomments{(1) in-image seeing of observations; (2) airmass of observations; (3) heliocentric distance of C/2025 K1; (4) geocentric distance of C/2025 K1; (5) phase angle; (6) geocentric and target orbital plane angle.}
\end{deluxetable*}

\begin{figure}
    \centering
    \includegraphics[width=0.8\linewidth]{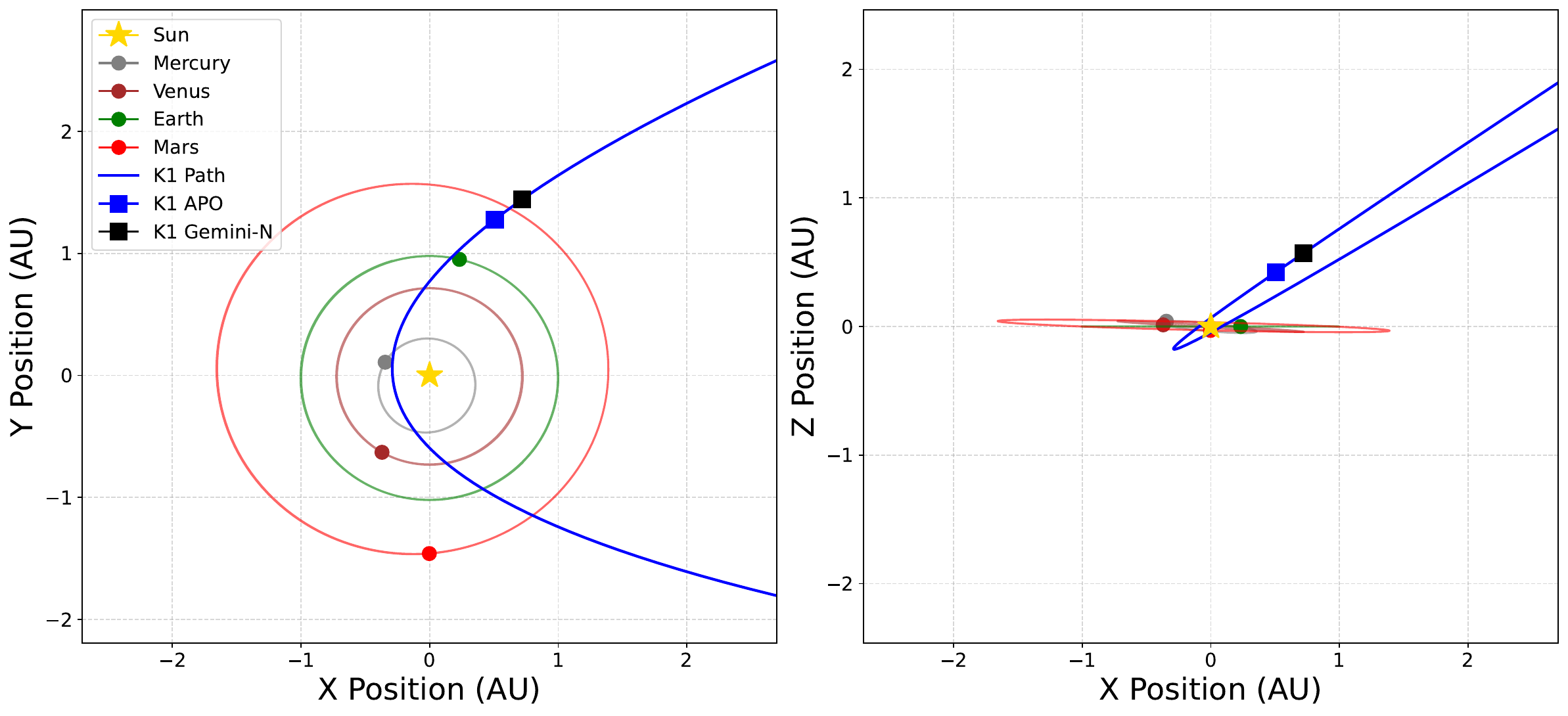}
    \caption{The position of C/2025 K1 is shown relative to the positions of the terrestrial planets during the on 2025 December 8 APO (blue) and 2025 December 24 Gemini North observations (black). All depicted orbits were obtained from the JPL Horizons ephemeris service. }
    \label{fig:orbit_plot}
\end{figure}

\begin{figure}
    \centering
    \includegraphics[width=1.0\linewidth]{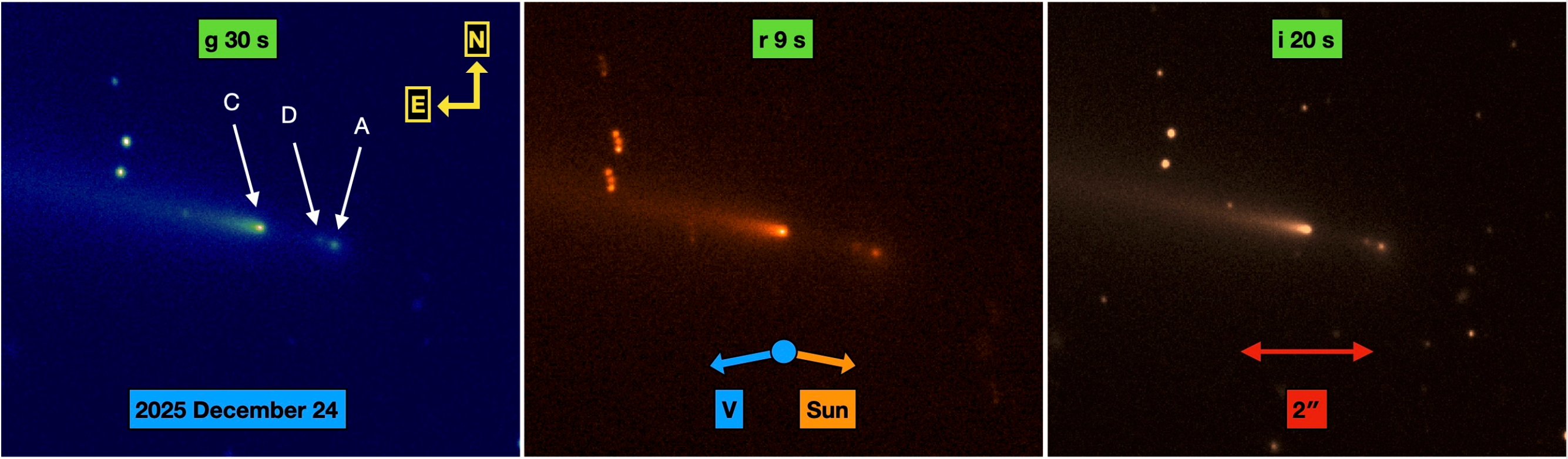}
    \caption{Gemini/GMOS g-, r-, and i-band data of C/2025 K1 taken on December 24, 2025. The individual fragments, cardinal direction, as well as, solar direction, heliocentric velocity direction, and image scale are indicated. A background star is present in the tail of the C fragment in the g-band image.}
    \label{fig:GemDec24}
\end{figure}

The second Gemini epoch was obtained on 2025 December 24 under program GN-2025B-DD-115 (PI: Bolin). The sequence extended from 06:17:27 to 06:20:37. The telescope again tracked at the non-sidereal rate of the comet, of $1\farcs59$~min$^{-1}$. We obtained one 30~s exposure through the g filter, three 3~s exposures through the r filter, and one 20~s exposure through the i filter. The nominal central wavelengths of these filters were 465~nm, 622~nm, and 771~nm, respectively. The seeing measured from the FWHM of in-frame background stars was approximately $0\farcs6$ in the r filter, and the observations covered an airmass range of 1.11--1.12. At the midpoint of the sequence, the heliocentric distance was 1.72~au, the geocentric distance was 1.06~au, and the phase angle was $31\fdg5$. The location of K1 during the 2025 December 24 Gemini observations is shown in Fig.~\ref{fig:orbit_plot}.  As seen in Fig.~\ref{fig:GemDec24}, Components A, D, and C were detected during the 2025 December 24 observations. Their projected separations from A were approximately 3\arcsec~for D and 13\arcsec~for C, corresponding to approximately 2300~km and 10,000~km at the distance of the comet, respectively. Component E, which was visible on December 6, was not detected on December 24, and the short observational arc makes predicted its location difficult. A dust tail $\sim$57\arcsec long was detected primarily in association with component C and had a position angle of approximately 78\degr.  Measured between the December 6 and December 24 epochs, fragments A and D were separating from each other by projected plane-of-sky rates of $\sim$1 m~s$^{1}$, and fragments A and C were separating from each other by projected plane-of-sky rates of $\sim$3~m~s$^{1}$, comparable to the separation speeds found for the K1 fragments observed in \textit{HST} images in 2025 November \citep[][]{Bodewits2026K1}.

We reduced the images using DRAGONS version 3.2.0 \citep[][]{gemini_dragons,Labrie2023DRAGONS,Labrie2023DRAGONSZenodo}. The processing included overscan correction, bias subtraction, flat-field correction, bad-pixel masking, cosmic-ray treatment, and i-filter fringe correction for the i-band image. Photometric calibration was performed using background stars found in Pan-STARRS \citep[][]{Chambers_2016_Panstarrs}. Fragment photometry was measured using a 19 pixel ($3\farcs1$) radius aperture in the 2025 December 6 GMOS data, and a 10 pixel ($1\farcs6$) radius in the 2025 December 24 data, which correspond to 1250~km radius at the distance of the comet at the time of the Gemini observations ($\sim$0.55~au on 2025 December 6 and $\sim$1.06~au on 2025 December 24). Astrometry of the comet and its fragments from the 2025 December data were measured and reported to the Minor Planet Center \citep[][]{Sato2025MPECK1Astro}.

\subsection{Apache Point Observatory}
\label{sec:apo}

\begin{figure}
    \centering
    \includegraphics[width=1.0\linewidth]{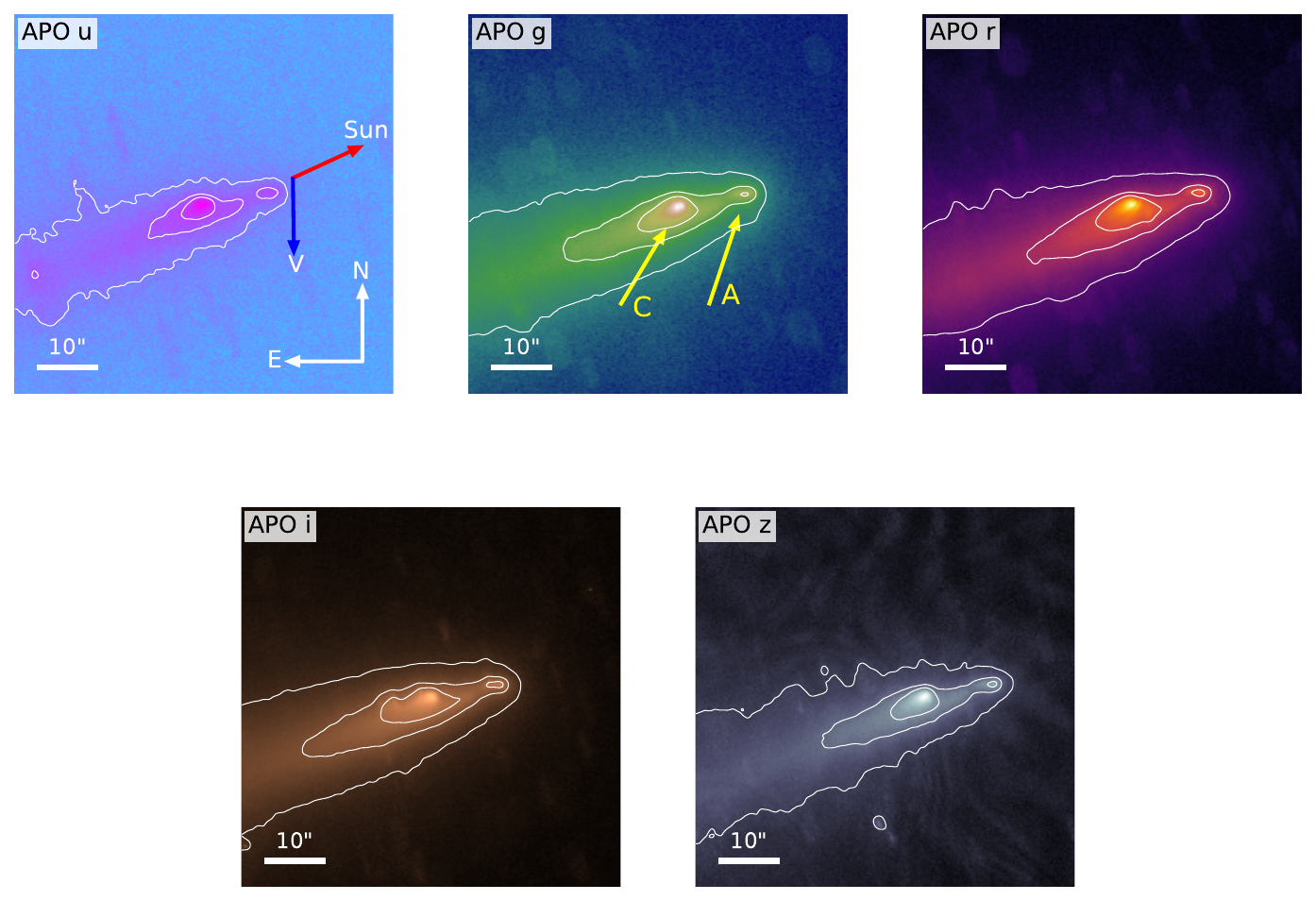}
    \caption{The APO observations of C/2025 K1 were taken in the five SDSS ugriz filters on the night of December 8, 2025. The median combined and stacked images of K1 in each filter are displayed here. The individual fragments, cardinal direction, as well as solar direction, heliocentric velocity direction, and image scale are indicated. Unresolved fragments A+D are labeled A in this figure. The two main fragments are visible in each filter, and photometry was performed on both of the fragments to measure the colors of both fragments.}
    \label{fig:APO_imaging}
\end{figure}

The observations of C/2025 K1 with the ARC 3.5-m telescope were obtained on 2025 December 8 from 05:50:32 to 08:20:03 UTC under program Q2JH04 (PI: Lisse). The heliocentric orbit and location of K1 at the time of the observations are shown in Figure~\ref{fig:orbit_plot}.  We imaged the comet with the Astrophysical Research Consortium Telescope Imaging Camera ARCTIC \citep{arctic_ref} through the five SDSS u, g, r, i, and z filters. The telescope tracked at the non-sidereal rate of the comet obtained from the Minor Planet Center ephemeris. The seeing was ($1\farcs22$), and the airmass ranged from 1.22 to 1.64 during the observing sequence. At the time the APO observations were taken, the heliocentric distance of K1 was 1.44~au, the geocentric distance was 0.59~au, and the phase angle was $31\fdg9$.  To minimize possible light-curve effects and filter-order-dependent biases, the filters were cycled in a randomized order. ARCTIC was operated with (2$\times$2) pixel binning, providing a pixel scale of $0\farcs228$~pixel$^{-1}$, and the observations followed a five-position dither pattern with (25\arcsec) offsets. The sequence comprised 6 u images, 8 g images, 21 r images, 5 i images, and 5 z images. The larger number of r images resulted from the use of that filter for target acquisition and placement of the comet within a favorable quadrant of the detector. Individual exposure times were 60~s in g, r, i, and z and 120~s in u, giving total integration times of 720, 480, 1260, 300, and 300~s in u, g, r, i, and z, respectively. As seen in Fig.~\ref{fig:APO_imaging}, the appearance of the comet and its fragments in the 2025 December 8 APO images was similar in appearance to the comet in the 2025 December 6 Gemini image, although fragments A and D are not separable due to the worse seeing of the APO images compared to the Gemini images.  At the end of the observing sequence, 21 bias frames and 5 sky flats in each filter were obtained. The observing geometry and additional circumstances are summarized in Table~\ref{tab:obs} and the heliocentric location of K1 at the time of the APO observations is shown in Fig.~\ref{fig:orbit_plot}.

We used standard image-reduction techniques to process the APO ARCTIC multi-band imagery \citep[e.g.,][]{Perley2025}. The median bias was subtracted from all images before the flats for each filter median-combined and normalized. All target images were divided by the normalized median flat for the matching filter. From the reduced images, we formed two sets of stacks: (1) a mean, median, and sum combined stack for each filter aligned on the centroid of the largest comet fragment (Fragment C), and (2) a mean, median, and sum combined stack aligned based on the background stars in the images.  Photometry was checked between both stacks before proceeding with the comet centered stack for all subsequent analysis.  We detect only two fragments -- A and C -- in these observations.  To enable direct comparison to the Gemini observations, we also adopted aperture sizes corresponding to the same $\sim$1250~km projected radius at K1's observer distance ($\sim$0.59~AU).  For these observations, a $2\farcs74$ radius corresponds to a physical size of $\sim$1250~km at the $\sim$0.59~au observer-distance of the comet.

We also used all calibrated images to calculate photometric zero points by comparing their magnitudes with those of stars in the Pan-STARRS catalog \citep{Chambers_2016_Panstarrs,flewelling2020panstarrs,Bolin2022IVO}. We used observed field-star magnitudes in the Pan-STARRS catalog to calculate effective photometric zero points in the calibrated images. We then used these zero points to calculate the magnitude of the comet fragments in each image. The comet magnitudes from each image were median-averaged to arrive at the magnitude we report. The errors for the median averaged magnitudes were obtained by computing the standard deviation of the individual frame measured magnitude values. The Pan-STARRS magnitudes were converted to SDSS magnitudes using the transformation equations in \citet[][]{Tonry2012}. Using the measured counts from the comet fragments, the background counts, and the photometric zero-points, we calculated the comet-fragment magnitudes in each filter for the mean, median, and sum images. Because the Pan-STARRS catalog lacks a u-filter, u-band photometry was obtained by comparing the comet to stars in the field with u-magnitudes from the SDSS catalog \citep{abdurro2022_DR17,fukugita1996sloan_filters}.

\section{Results} 
\label{sec:analysis}

\subsection{Photometry and Multi-Band Colors}
\label{sec:colors}

The magnitudes of the comet fragments in each filter from Gemini North and APO are summarized in Table~\ref{tab:magnitudes}. For each detected condensation, we subtracted the magnitudes measured in adjacent SDSS filters to determine the colors listed in Table~\ref{tab:colors}. For the Gemini observations, we find $g-r=0.34\pm0.03$, $0.42\pm0.04$, and $0.32\pm0.04$ for fragments C, D, and A, respectively, and $r-i=0.20\pm0.03$, $0.22\pm0.04$, and $0.30\pm0.04$, respectively. For the APO observations, fragment C has $u-g=1.52\pm0.21$, $g-r=0.35\pm0.09$, $r-i=0.17\pm0.07$, and $i-z=-0.01\pm0.07$. The blended A+D condensation has $u-g=1.58\pm0.36$, $g-r=0.35\pm0.08$, $r-i=0.17\pm0.08$, and $i-z=+0.02\pm0.10$. Within the uncertainties, the fragment colors measured by Gemini North and APO are consistent with each other.  Since these observations were taken $\sim16$~days apart, we have confidence that these results are not spurious or influence by instrumental uncertainty.

\begin{deluxetable}{llccccc}
\tablecaption{APO and Gemini Fragment Magnitudes
\label{tab:magnitudes}}
\tablehead{
\colhead{Observatory} &
\colhead{Fragment} &
\colhead{$u$} &
\colhead{$g$} &
\colhead{$r$} &
\colhead{$i$} &
\colhead{$z$}
}
\startdata
APO & C & $17.40\pm0.20$ & $15.88\pm0.07$ & $15.53\pm0.05$ & $15.36\pm0.05$ & $15.37\pm0.05$ \\
        & A & $18.35\pm0.36$ & $16.77\pm0.06$ & $16.42\pm0.06$ & $16.25\pm0.06$ & $16.23\pm0.08$ \\
\hline
       Gemini & C & \nodata & $18.34\pm0.03$ & $18.00\pm0.01$ & $17.80\pm0.03$ & \nodata \\
        & D & \nodata & $19.89\pm0.03$ & $19.47\pm0.02$ & $19.25\pm0.03$ & \nodata \\
        & A & \nodata & $19.52\pm0.03$ & $19.20\pm0.02$ & $18.90\pm0.03$ & \nodata \\
\enddata
\tablecomments{Fragments A and D are unresolved from one
another in the APO images; their blended photometry is therefore
reported as A+D.}
\end{deluxetable}

\begin{deluxetable}{llcccc}
\tablecaption{Broadband Colors of the C/2025 K1 Fragments
\label{tab:colors}}
\tablehead{
\colhead{Observatory} &
\colhead{Fragment} &
\colhead{$u-g$} &
\colhead{$g-r$} &
\colhead{$r-i$} &
\colhead{$i-z$}
}
\startdata
APO    & C   & $1.52\pm0.21$ & $0.35\pm0.09$ & $0.17\pm0.07$ & $-0.01\pm0.07$ \\
    & A+D & $1.58\pm0.36$ & $0.35\pm0.08$ & $0.17\pm0.08$ & $+0.02\pm0.10$ \\
    \hline
Gemini & C   & \nodata       & $0.34\pm0.03$ & $0.20\pm0.03$ & \nodata \\
 & D   & \nodata       & $0.42\pm0.04$ & $0.22\pm0.04$ & \nodata \\
 & A   & \nodata       & $0.32\pm0.04$ & $0.30\pm0.04$ & \nodata \\
\enddata
\tablecomments{Colors were calculated by subtracting the magnitudes
reported in Table~\ref{tab:magnitudes}. Their uncertainties were
propagated in quadrature. The APO A+D measurement represents the
blended condensation because fragments A and D are unresolved in the
APO images.}
\end{deluxetable}

\begin{figure}
    \centering
    \includegraphics[width=0.8\linewidth]{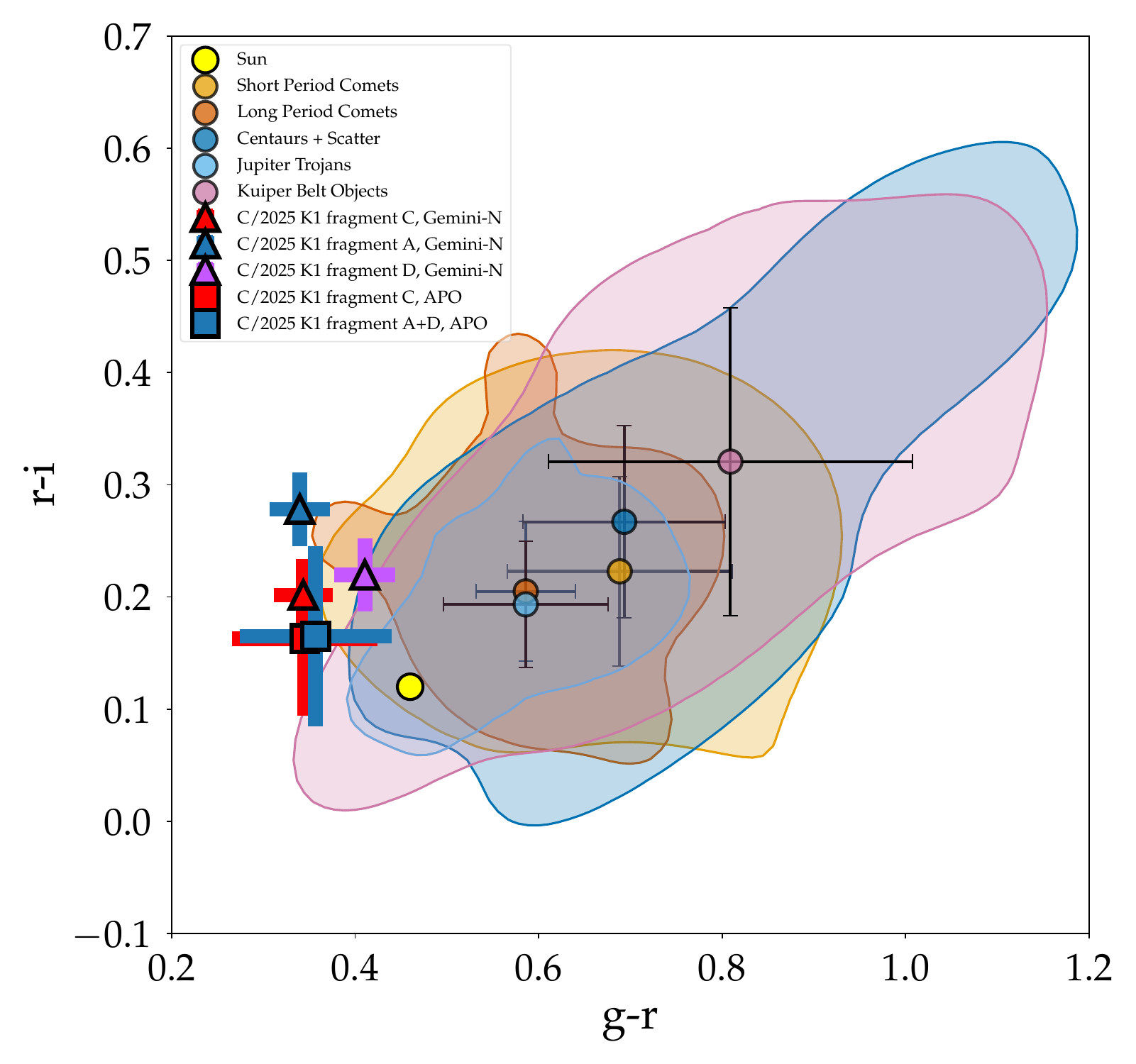}
    \caption{The SDSS g-r and r-i photometric colors of C/2025 K1 ATLAS observed by Gemini and APO are consistent with each other, as seen in this plot inspired by Jewitt \citep{jewitt2015color}. The Gemini observations are from Dec. 24, 2025, and the APO observations are from Dec. 8, 2025. Three fragments were measured in the Gemini data, while only the two largest were visible in the APO imaging. The colored contours are the regions of color-color space occupied by the different families of solar system objects. The contours have been determined using kernel density estimation to map the color space occupied by each population. The median g-r and r-i colors, with 1$\sigma$ error bars calculated from the median absolute deviation, for each solar system minor body family have also been plotted. All of the solar system photometry to calculate the colors was retrieved from the MBOSS dataset \citep{MBOSS_2012}.}
    \label{fig:K1_color_plot}
\end{figure}

Using the measured colors of the comet fragments, we then constructed color-color plots to compare with solar system comets, Jupiter Trojans, and Kuiper Belt Objects (KBOs) from the MBOSS catalog \citep[e.g.,][]{MBOSS_2012,Bolin2023NTs}. The Gemini and APO $g-r$ and $r-i$ colors are plotted together in Fig.~\ref{fig:K1_color_plot}. Overall, we find that the $g-r$ photometric colors of K1 are bluer than solar system comparisons. Notably, the colors do not lie on the reddening curve for solar system objects constructed by \cite{jewitt2015color}. However, there are some previously known comets that do have similarly blue colors. Although the central $u-g$ colors for the A and C fragment measured in the APO data are slightly redder than the solar $u-g=1.43$ \citep[][]{Willmer2018}, neither fragment's difference is significant outside of the uncertainties. The lack of a detection of fragment B in the 2025 December 6 Gemini seen in previous images of the comet \citep[][]{Kostov2025K1Fragment,Guzik2025FragmentK1} could be due to the dissipation of the fragment or fading below the sensitivity limit of our observations. We find a 3-$\sigma$ lower limit on the magnitude of Fragment B (location in Fig.~\ref{fig:GemDec6} labeled B) in SDSS r-band of $\sim$18.9.

\begin{figure}
    \centering
    \includegraphics[width=0.80\linewidth]{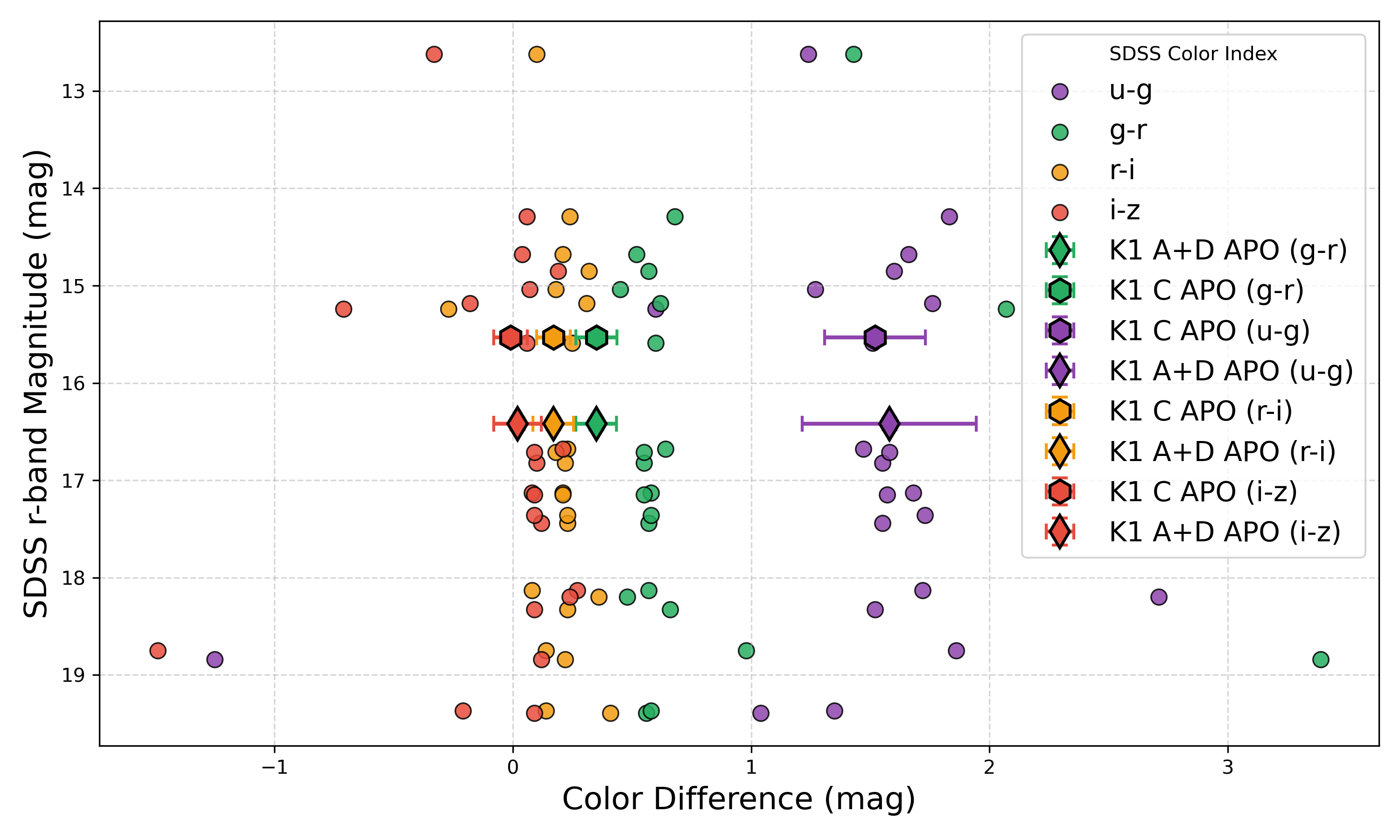}
    \caption{The colors calculated from the APO observations in the five SDSS filters where compared to the colors of a population of 31 comets observed by the SDSS survey \citep{Solontoi_2012} From this comparison we find that the u-g, r-i, and i-z colors of the two largest K1 fragments are consistent with the bulk of the comet population observed by SDSS. Only the g-r color is discrepant.}
    \label{fig:APO_Color_Comparison}
\end{figure}

From the SDSS r-filter images taken with both Gemini North and APO, we also calculated the A\textit{f}$\rho$ parameter for each fragment using the definition from \citet{AHearn1984afrho} and \citet{AHearn_1995}:
\begin{align}
    A\textit{f}\rho=\frac{4r_\text{hel}^2\Delta^2}{\rho_{ap}}10^{-0.4(m_\text{comet}-m_{\odot})},
\end{align}
where $r_\text{hel}$ is the heliocentric distance in AU, $\Delta$ is the geocentric distance to the comet in cm, $\rho_{ap}$ is the projected aperture radius used in the photometry in cm, and $m_\text{comet}$ and $m_{\odot}$ are the magnitudes of the comet and Sun in the SDSS filters, respectively \citep{Fink2012}. Since $\rho_{ap}$ denotes the aperture radius, the numerator contains the standard factor of 4. We used the same aperture radius to calculate A\textit{f}$\rho$ for the Gemini North and APO observations as we did when measuring the photometry which we list in Table~\ref{tab:afrho}. We caution that the assumptions for using A\textit{f}$\rho$ as a aperture-independent proxy for dust production may be limited in the case of C/2025 K1. As discussed in \citet{Fink2012}, A\textit{f}$\rho$ assumes that the projected dust column density follows a 1/$\rho$ profile, spherical symmetry, and constant steady-state outflow. \citet{Bodewits2026K1} observed C/2025 K1 to have experienced multiple outburst events concurrent with its fragmentation, which increased dust production over a short time period and was associated with the evolution of an asymmetric coma. Residual outburst dust and the overlap of different fragments' comae may provide additional difficulties for the application of A\textit{f}$\rho$ as an activity indicator. With these caveats in mind, we interpret A\textit{f}$\rho$ as an empirical estimate of coma brightness within the specified aperture, which is still useful to compare between observations taken at different epochs as an indicator of the evolution in the comet's activity.

The December 6th observations were taken at a phase angle of $32\fdg4$, the December 8th observations at a phase angle $31\fdg9$ and the December 24th observations at a phase angle $31\fdg5$. With this information, we calculated the phase-corrected A\textit{f}$\rho$($0^\circ$) values for each observation. We used the low-phase-angle Halley-Marcus composite phase function \citep[][]{Schleicher1998,Marcus2007,Schleicher2011}:
\begin{equation}
\Phi_{\rm HM}(\alpha)
= 10^{-0.01807\alpha + 0.000177\alpha^2},
\label{eq:hm_phase}
\end{equation}
which is described by \citet{Marcus2007} and \citet{Schleicher2011} to be applicable for $0^\circ \leq \alpha \leq 42^\circ$ and $\Phi_{\rm HM}(0^\circ)=1$. All of our observations of C/2025 K1 lie within this range. Then, the phase-corrected A\textit{f}$\rho$ values are defined as:
\begin{equation}
Af\rho(0^\circ)
= \frac{Af\rho(\alpha)}{\Phi_{\rm HM}(\alpha)}.
\label{eq:afrho_phase_corrected}
\end{equation}
We note that the uncertainty on A\textit{f}$\rho$($0^\circ$) may be dominated by systematics in adopting this phase function which may not precisely represent the true phase function of C/2025 K1's grains. We use this equation to calculate A\textit{f}$\rho$($0^\circ$) for fragments A and C for all three nights (with A and D unresolved in the APO images) and for fragments D and E when possible, as listed in Table~\ref{tab:afrho}.

\begin{deluxetable}{llcc}
\tablecaption{A\textit{f}$\rho$ values for C/2025 K1
\label{tab:afrho}}
\tablehead{
\colhead{Observatory} &
\colhead{Fragment} &
\colhead{A\textit{f}$\rho$ [cm]} &
\colhead{A\textit{f}$\rho$($0^\circ$) [cm]}
}
\startdata
Gemini 6 December & A & $24.70\pm1.00$ & $62.00\pm2.51$  \\
        & C & $56.66\pm1.00$ & $142.21\pm2.51$ \\
        & D &  $24.32\pm1.00$ & $61.04\pm2.51$ \\
        & E & $29.14\pm1.00$ & $73.14\pm2.51$ \\
        \hline
        APO 8 December & A+D & $25.30\pm2.60$ & $63.01\pm6.48$ \\
        & C & $57.16\pm4.90$ & $142.37\pm12.20$ \\
        \hline
        Gemini 24 December & A & $11.86\pm2.00$ & $29.35\pm4.95$  \\
        & C & $29.48\pm2.00$ & $72.96\pm4.95$ \\
        & D & $3.54\pm2.00$ & $8.76\pm4.95$ \\
\enddata
\tablecomments{A and D are unresolved in the APO images, so their 8 December value is reported for the blended A+D condensation. The phase-corrected values use the Halley--Marcus phase function in Equation~\ref{eq:hm_phase}. The quoted uncertainties propagate the statistical uncertainties in A\textit{f}$\rho$ and do not include systematic uncertainty in the adopted phase function.}
\end{deluxetable}

\subsection{Dust Ejection Velocity Calculations}

From the Gemini North r-band imaging of C/2025 K1, we determined the dust ejection velocity. We chose r-band, because it has high signal-to-noise and this wavelength is dominated by light scattering by dust, and measured the extent of the dust coma along the comet-Sun position vector. Unfortunately, in the APO r-band data, overlap from the comet fragments made it impossible to measure the extent of the coma cleanly in the sunward direction. We employed the `fountain-model' of dust coma, assuming that the dust is emitted with some velocity and is eventually stopped and pushed back by solar radiation pressure \citep[][]{jewitt1987}. The dust ejection velocity is given by:
\begin{align}
    v_{ejection}=\frac{(2D_t\beta g_{\odot})^{\frac{1}{2}}}{r[\text{au}]},
\end{align}
where $D_t$ is the distance of the coma in the sunward direction in kilometers, $g_{\odot}$ = 0.006~m~s$^{-2}$ is the acceleration due to the Sun's gravity at 1~au, and $r$ is the distance of the comet from the Sun in au \citep[listed in Table~\ref{tab:obs} for each date of observations,][]{Hsieh2021QN173,Bolin20253I}. $\beta$ is defined as the ratio of the acceleration of a grain due to radiation pressure relative to the local solar gravity:
\begin{align}
\frac{5.7\times10^{-4} Q_{pr}}{\rho_\text{dust}r_\text{dust}},
\end{align}
where $\rho_\text{dust}$ is the mass density of dust, $r_\text{dust}$ is the radius of the dust, and $Q_{pr}$ is the radiation pressure efficiency which we took to be $\sim1$ for grains much larger than the wavelength of light. For millimeter-sized dust, we used a value of $\beta=0.00076$, and for micron-sized dust we used $\beta=1.0$. Table~\ref{tab:eject} presents the results of this calculation for each fragment of C/2025 K1 for which we measured an $Af\rho$ value and have tabulated these ejection speeds.

\begin{deluxetable}{llcc}
\tablecaption{Dust Ejection Velocities for the Fragments of C/2025 K1
\label{tab:eject}}
\tablehead{
\colhead{Observatory} &
\colhead{Fragment} &
\colhead{$v_{\rm ej}(a=1\,{\rm mm})$} &
\colhead{$v_{\rm ej}(a=1\,\mu{\rm m})$} \\
\colhead{} &
\colhead{} &
\colhead{[m\,s$^{-1}$]} &
\colhead{[m\,s$^{-1}$]}
}
\startdata
Gemini 6 December & A & $2.41 \pm 0.33$ & $76.19 \pm 10.97$ \\
                 & C & $2.26 \pm 0.29$ & $71.31 \pm 11.69$ \\
                 & D & \nodata & \nodata \\
    \hline
    Gemini 24 December & A & $3.59 \pm 0.23$ & $98.92 \pm 6.97$ \\
                  & C & $3.70 \pm 0.22$ & $101.89 \pm 6.67$ \\
                  & D & $1.53 \pm 0.36$ & $41.50 \pm 10.81$ \\
\enddata
\tablecomments{Dust ejection velocities were calculated
        from the Gemini imaging obtained on 6 and 24 December for
        assumed grain radii of 1~mm and 1~$\mu$m. Source confusion
        among the fragments prevented an estimate of the dust
        ejection velocity for fragment D on 6 December.}
\end{deluxetable}

\section{Discussion} 
\label{sec:discussion}

We have presented new photometric observations of fragmenting comet C/2025 K1 with both Gemini North and APO.  These new multi-color observations provide an important new data point following the comet's close perihelion passage and subsequent fragmentation. Here, we discuss the fragment colors in the context of other solar system bodies and previously observed fragmenting comets.

\subsection{Color Comparison to Other Solar System Objects}
\label{sec:colorcomparison}

All of the C/2025 K1 fragments have blue g-r colors compared to other Solar System bodies.  In Figure~\ref{fig:K1_color_plot}, we specifically compare with Short Period Comets, Long Period Comets, Centaurs, Jupiter Trojans, and KBOs.  The shaded regions show the spread in measurements with the points giving the median values.  Although the medians are quite far from the measured K1 fragment colors, some of the shaded regions overlap within the uncertainties.  Unlike these representative populations, however, Comet C/2025 K1 is dynamically new and likely originates from the Oort Cloud.  Thus, it is perhaps not surprising that its color is somewhat different. 

Other previously observed comets have shown similar blue colors.  Comet C/2016 R2 PanSTARRS exhibited strong gas emission (CO$^+$ and N$_2^+$) along with a corresponding blue $g-r = 0.40\pm0.05$ \citep{Biver:2018,McKay:2019}.  Comet C/2017 K2 PanSTARRS exhibited $g-r$ colors ranging from $0.30-0.45$ \cite[e.g.,][]{Hmiddouch:2025}. Apparent blue colors can result from strong gas emission lines overlapping within broadband g-filters such as in the case of interstellar comet 3I/ATLAS \citep[e.g., CN, C$_2$, CO$^+$,][]{Rahatgaonkar20253IVLTgas}.  However, both these comets were distantly active and volatile-rich, unlike C/2025 K1.  Additionally, the $g-r$ and $r-i$ colors of fragments A and C are consistent between 2025 December 8 and 24 within the reported photometric uncertainties, with no statistically significant color change detected over this interval. Both epochs sample the post-fragmentation system; these observations therefore do not establish whether the fragment colors differ from those of the comet before fragmentation.

Comparisons between the colors of active objects such as K1 and inactive small bodies require caution because the measurements sample different scattering environments. K1's measured colors predominantly characterize the comae surrounding its fragments, whereas observations of inactive objects characterize their solid surfaces. As discussed in \citet{Zubko2020}, these differences can produce a blue coma and a red surface without requiring different compositions. Consequently, comparisons with the surface colors of inactive Centaurs, Trojans, and KBOs provide observational context but do not independently establish compositional similarities or differences, or evidence for freshly exposed ice.

Additionally, we compare the APO five-band SDSS photometry to the population of comets observed by SDSS presented in Solonoi et al in \ref{fig:APO_Color_Comparison} \citep{Solontoi_2012}. This set of observations consisted of 31 comets observed in the five SDSS filters with APO, making it the closest available data set to compare with our K1 photometry. We perform a two-tailed z-test for each of the four colors and calculate a multi-variate chi-squared test with all four colors. We find that the color deviations for each of the four colors is -0.2-$\sigma$ for u-g, -1.5-$\sigma$ for g-r, -0.2-$\sigma$ for r-i, and 0.2-$\sigma$ for i-z. The result of the multi-variate chi-squared test is $\chi^2=2.47$ with a p-value of $p=0.65$. The bluer g-r color is suggestive of possible gas emission in the SDSS g-band, but the multi-variate chi-squared test indicates that the K1 colors are not significantly different from the comet population observed by SDSS as a whole.

Ultraviolet photometry of comets has been used to measure OH production by observing near the OH (0-0) band at 309~nm, which can serve as a proxy for water production \citep{AHearn_1995_phot}. Ground-based broadband u-band photometry, such as that from SDSS and Rubin/LSST, is only very weakly sensitive to OH because of atmospheric absorption at these wavelengths. However, it does sample the near-UV continuum along with contributions from gas species such as NH and CN \citep{Farnham2000}. Large samples of u-band comet photometry could therefore constrain the near-UV colors of cometary dust and the level of gas contamination in broadband measurements, complementing dedicated narrow-band and space-based observations.

\subsection{Fragmenting Comets}
\label{sec:fragmentingcomets}

Some periodic comets can experience unusually large dust outbursts, such as 17P/Holmes in its 2007 apparition \citep{schleicher2009_17P, dello2008_17P}.  This event produced a large dust shell that expanded around the nucleus, causing a large increase in brightness followed by a dramatic decline. The cause of this outburst is still not fully understood, but it could have resulted from the shedding of the outer layer of the comet's surface. When a comet fragments or has a large dust ejection event, large amounts of small dust grains are ejected that dramatically increase the surface area of the comet, such as in the cases of active Main Belt asteroids  P/2016 J1 and (6478) Gault \citep[][]{Moreno2017J1,YeGault2019}. This larger surface area both allows more light to be reflected and more volatiles to be exposed so that they can sublimate.  Comet C/2025 K1 could have experienced an analogous outburst during fragmentation if the removal of the outer surface layer exposed a layer of volatiles observed spectroscopically in November 2025. However, after the dust and gas are dispersed by solar radiation, the comet returns to a quiescent state as observed it in December 2025.

Comet 73P/Schwassmann-Wachmann has been observed to be disintegrating since 1995. The fragmentation became especially dramatic when 73P approached perihelion in 2006 with the comet fragmenting into more than 66 pieces \citep{Reach_2009_73P,Graykowski2019}. Spectroscopic observations of 73P demonstrated that it was also volatile-depleted, especially in carbon-chain species such as ethane, carbon monoxide, and methanol relative to water \citep{Kobayashi_2007_73P,Villanueva_2006_73P}. Interestingly, hydrogen cyanide was shown to still have a normal abundance relative to water \citep{russo2007_73P}.  Overall, the comet appeared chemically homogeneous. The mineralogy of 73P was dominated by amorphous carbon and amorphous pyroxene, demonstrating that the carbon present was not in the gas-phase species \citep{harker2011_73P}. Given these results, 73P has been interpreted as having a volatile-poor primordial composition, reflecting a formation location closer to the proto-Sun in the solar system's protoplanetary disk \citep{russo2007_73P}. Like 73P, K1 fragmented into multiple components and exhibits depletion in some carbon-bearing species.  However, their different dynamical histories and the limited abundance measurements for K1 do not establish a common formation environment.

\section{Conclusions} 
\label{sec:conclusions}

In this paper, we analyze new multi-band photometric imaging of fragmenting comet C/2025 K1 spanning 2025 December 6 through 24.  Our main conclusions are:

\begin{enumerate}
    \item The photometry of the comet C/2025 K1 is blue relative to many families of solar system small bodies (Fig.~\ref{fig:K1_color_plot}). The photometric $g-r$ and $r-i$ colors in SDSS filters for the A and C fragments -- $g-r\simeq0.35$ and $r-i\simeq0.2$ -- are consistent with each other and between the APO and Gemini North photometry. Since we made these measurements $\sim$16 days apart, this gives confidence that our photometric results are not spurious.

    \item We calculated the A\textit{f}$\rho$($0^\circ$) parameter from the SDSS r-band images obtained with Gemini North and APO. Since the fragmented and asymmetric comae of C/2025 K1 may not satisfy the assumptions underlying A\textit{f}$\rho$, we interpret these measurements as empirical indicators of coma brightness rather than direct dust-production rates. For fragment A, A\textit{f}$\rho$($0^\circ$) was $62.00\pm2.51$~cm on December 6 and $29.35\pm4.95$~cm on December 24.  The unresolved A+D condensation measured $63.01\pm6.48$~cm on December 8. Fragment C showed a clear decrease in A\textit{f}$\rho$($0^\circ$) from $142.21\pm2.51$ cm on December 6 and $142.37\pm12.20$~cm on December 8 to $72.96\pm4.95$~cm on December 24.
    
\end{enumerate}

\noindent The time-resolved observations of C/2025 K1, give us a glimpse into how its dust population has evolved since the initial fragmentation event in 2025 November.  In the future, the Vera C. Rubin Observatory will likely detect many more fragmenting objects allowing for similar detailed follow-up \citep{Schwamb2023LSST,Andreoni2024RubinToO}.  By building a comparison population, we can gain new insights into the formation history of our planetary system.

\begin{acknowledgments}

The authors wish to thank the editor and the anonymous referee for their constructive feedback which was crucial in the improvement of the manuscript.

This work is based in part on observations obtained at Apache Point Observatory 3.5 m telescope, which is owned and operated by the Astrophysical Research Consortium, in the Sacramento Mountains of southern New Mexico. We acknowledge this region as part of the traditional homelands of the Mescalero Apache people and recognize their enduring stewardship and cultural relationship to this landscape.

This work is based in part on observations obtained at the Gemini North telescope on Maunakea, Hawai'i. We acknowledge Maunakea as a place of deep cultural and spiritual importance to Kanaka'oiwi, the Native Hawaiian community, and we recognize the privilege and responsibility of conducting astronomical research from this sacred summit.

This research benefited from B. T. Bolin's participation in the 2026 workshop ``Exploring Planetary Systems in the Era of Time-domain Astronomy'', hosted by the Institute for Astronomy at the University of Hawaii and supported by award 2106927 from the National Science Foundation's Astronomy \& Astrophysics Research Grants program.

Part of this research was conducted at the Jet Propulsion Laboratory, California Institute of Technology, under a contract with the National Aeronautics and Space Administration (80NM0018D0004)

\end{acknowledgments}

\begin{contribution}

Carl Ingebretsen (C.I.) as the lead author, was primarily responsible for the drafting of this paper. He was primarily responsible for the data collection from APO on December 8, and the reduction and analysis of the APO data. He also made the photometry measurements for the Gemini North Data. C.I. was responsible for submitting the manuscript

Bryce Bolin was the principal investigator for the Gemini North observations of K1. He reduced the data from these observations and contributed to the text, figures, and the revision of the draft. 

Meredith MacGregor was responsible for drafting portions of the manuscript and advising C.I. on the draft and interpretation of the results.

Carey Lisse was the principal investigator for the APO observations. He assisted in the data collection and advised C.I. on the interpretation of the results.

Matthew Belyakov assisted in the APO observations and contributed code to help C.I. analyze the data.

Gracyn Jewett assisted in the reduction of the APO data.

Mukremin Kilic assisted in the observations of the comet with APO.

Marco Micheli measured the astrometry of the comet and its fragments in the Gemini data.

Davide Farnocchia checked the astrometry of the comet fragments and computed their orbits.

Brian Lemaux provided support and exposure design for the Gemini observations.

Hyewon Suh helped support and execute the Gemini observations.


\end{contribution}

%
\facilities{Facilities: Gemini:North (GMOS-N), ARC (ARCTIC)}

\software{astropy \citep{2013A&A...558A..33A,2018AJ....156..123A,2022ApJ...935..167A},
IRAF \citep{iraf},
DRAGONS \citep{gemini_dragons,Labrie2023}}

\bibliographystyle{aasjournalv7.1}
\bibliography{references.bib}



\end{document}